\documentclass[showpacs,prd,showkeys,aps]{revtex4}

\usepackage{amsmath}
\usepackage{amssymb}

\usepackage{braket}
\usepackage{eurosym}
\usepackage{amsfonts}
\usepackage{graphicx}
\usepackage{calrsfs}
\usepackage[usenames,dvipsnames,svgnames,table]{xcolor}
\usepackage{hyperref}

\begin{document}

\title{Canonical Acoustic Thin-Shell Wormholes }

\author{Kimet Jusufi}
\email{kimet.jusufi@unite.edu.mk}

\author{Ali \"{O}vg\"{u}n}
\email{ali.ovgun@emu.edu.tr}

\affiliation{Physics Department, State University of Tetovo, Ilinden Street nn,
1200, Macedonia}

\affiliation{Physics Department, Eastern Mediterranean University, Famagusta,
Northern Cyprus, Mersin 10, Turkey}

\date{\today }

\begin{abstract}
In this paper we model a canonical acoustic thin shell wormhole (CATSW) in the framework of analogue gravity systems. In this model we apply cut and paste technique to join together two spherically symmetric, analogue canonical acoustic solutions, and compute the analogue surface density/surface pressure of the fluid using the Darmois\textendash Israel formalism. We study the stability analyses by using a linear barotropic fluid (LBF), chaplygin fluid (CF), logarithmic fluid (LogF), polytropic fluid (PF), and finally Van der Waals Quintessence (VDWQ). We show that a kind of analog acoustic fluid with negative energy is required at the throat to keep the wormhole stable. It is argued that, CATSW can be a stabile thin-shell wormhole if we choose a suitable parameter values.

\end{abstract}


\maketitle
 
\section{Introduction}

According to the General Theory of Relativity, there is a possibility to form a wormhole  by connecting two different spacetime regions of the universe \cite{fl,er}.  Recently traversable  wormhole solutions were found by Morris and Thorne \cite{mth1,mth2}, who argued that, if traversable wormholes exist, one needs to invoke a negative energy at the throat which sometimes is known as exotic matter \cite{mth2}. The idea was put forward by Visser, who introduced the concept of thin shell wormholes, which allows us to construct a wormhole simply by cutting and pasting two space-time regions \cite{mth2,mv1,mv2,mv3,is}. This method attracted a lot of interest among physicists and a number of papers have been written on this topic (see for example\cite{eiroa1,ayan1,lobo1,eiroa2,ali1,rah,mazh1,ali,kimet,po,eiroa3,ali2,ern1,ern2,lemos1,lemos2,ayan2,myrzakulov,kim,peter1,sharifcy1,sharifcy2,sh1,sh2,sh3}).

In a seminal paper, Unruh \cite{unruh1}, for the first time found an analogy connecting black holes and sonic black holes and raised the question whether is possible to detect the Hawking radiation (sometimes known as Hawking\textendash Unruh radiation) in a laboratory \cite{Jeff}. One can show that, under linearized perturbations, the equations of motion of a moving fluid is an equivalent way to describe the propagation of a massless scalar field in a curved spacetime \cite{visser1}. One particular example is to study the wave equation for sound waves in a moving fluid as an analogue for light waves in a curved spacetime\cite{carlos}.

In the context of analogue black hole models a number of interesting analogue metrics have been found, up to a conformal factor. In particular, the Schwarzschild type metric solution known as a canonical acoustic metric and the Painleve\textendash Gullstrand acoustic metric \cite{visser1}, a rotating analogue metric \cite{visser1,unruh2,visser2}, analogue AdS and dS black hole solutions \cite{lib} and references therein. Furthermore Nandi et al \cite{nandi} introduced the idea of acoustic traversable wormholes and showed that this analogy model can be used to investigate the nature of curvature singularity, to study the light ray trajectories in an optical medium is equivalent to the sound trajectories in its acoustic analog. On the other hand, acoustic black holes are studied to calculate the quasinormal modes, superradiance and area spectrum by Saavedra \cite{saavedra1,saavedra2}.

In this paper, we aim to model a spherically symmetric, thin shell wormhole within the framework of analogue gravity systems. In particular, we aim to explore the stability of the acoustic wormhole by modeling the analogue fluid with a linear barotropic fluid(LBF) \cite{kuf2,varela}, chaplygin fluid (CF) \cite{cg1,cg2,GCCG,GCCG1}, logarithmic fluid (LogF) \cite{ali1,ali2}, polytropic eqiation of state for the fluid, \cite{sarkar}, Van der Waals Quintessence (VDWQ) \cite{Van}. 

The paper is organized as follows.  In Section 2, we review the canonical acoustic metric. In section 3, we construct an analogue, canonical thin shell wormhole and show show that an analog exotic fluid with negative energy is required at the throat to keep the wormhole stable. In Section 4 we investigate the stability analyses by modeling the analogue fluid by a linear barotropic fluid (LBF), chaplygin fluid (CF), and logarithmic fluid (LogF), polytropic fluid and Van der Waals Quintessence (VDWQ). In Section 5, we comment on our results.

\section{Canonical Acoustic Black Holes}

We can start by writing the spherically symmetric solution of incompressible fluid or the so--called canonical acoustic black hole metric found by Visser written as follows \cite{visser1,vieira}
\begin{equation}
\mathrm{d}s^{2}=-c^{2}\mathrm{d}t^{2}_{l}+\left(\mathrm{d}r\pm c \frac{r_0^{2}}{r^2} \,\mathrm{d}t_{l} \right)^2+r^2 \left(\mathrm{d}\theta^2 +\sin^{2}\theta \mathrm{d}\phi^2 \right).\label{1}
\end{equation}

In which $c$ is the speed of sound through the fluid, $v$ is the fluid velocity, and $t_{l}$ is the laboratory time.  Morover $c$ is related with the velocity $v$ as follows
\begin{equation}
v=c \frac{r_0^2}{r^2}.
\end{equation}

However it is more convenient to write the metric \eqref{1} as a Schwarzschild type metric. If we introduce the Schwarzschild time coordinate $t$, which is related to the laboratory time $t_{l}$ by the following coordinate transformation
\begin{equation}
\mathrm{d}t\to \mathrm{d}t_{l} \pm \frac{r_0^2/r^2}{c \left(1-(r_0^4/r^4)\right)}\mathrm{d}r.
\end{equation}

Then, it's not difficult to show that we end up with a spherically symmetric acustic metric also known as a canonical acoustic black hole
as \cite{vieira,ramon}
\begin{equation}
\mathrm{d}s^{2}=-c^{2}f(r)\mathrm{d}t^{2}+\frac{\mathrm{d}r^{2}}{f(r)}+r^{2}(\mathrm{d}\theta^{2}+\sin^{2}\theta \,\mathrm{d}\phi^{2}) \label{4}
\end{equation}
in which 
\begin{equation}
f(r)=1-\frac{r_0^{4}}{r^{4}}.
\end{equation}

The event horizon of the canonical
acoustic black hole is computed by solving $g_{rr}(r_h) = 0$, so that one gets
\begin{equation}
r_{h}=r_0.
\end{equation}

The corresponding gravitational acceleration for the acoustic black hole horizon can be computed as follows
\begin{equation}
\kappa_{0}=\left.\frac{\partial v^r}{\partial r}\right\vert
_{r=r_{h}}=\frac{2 c}{r_0}.
\end{equation}

The analogue Hawking temperature for the canonical acoustic black holes due to the emission of phonons is given by  \cite{vieira}
\begin{equation}
T_{H}=\frac{\kappa_0}{2 \pi c}=\frac{1}{\pi r_0}
\end{equation}

Without loss of generality from now on we can set the speed of sound to unity, i.e. $c=1$. In what follows we are going to make use of the metric \eqref{4} to model CATSW.

\section{Canonical Acoustic Thin shell Wormholes }

Let us now proceed to use cut and paste technique to construct a CATSW using the
metric \eqref{4} and choosing two identical regions 
\begin{equation}
M^{(\pm)}=\left\lbrace r^{(\pm)}\geq a,\,\,a>r_{h}\right\rbrace ,
\end{equation}
in which $a$ is choosen to be grater than the event horizon $r_{h}$. If we now paste these regular regions at the boundary hypersurface $\Sigma^{(\pm)}=\left\lbrace r^{(\pm)}=a,a>r_{H}\right\rbrace $, then we end up with a complete manifold $M=M^{+}\bigcup M^{-}$.
In accordance with the Darmois\textendash Israel formalism the coordinates on $M$ can be choosen as $x^{\alpha}=(t,r,\theta,\phi)$. On the other hand for the coordinates on the induced metric $\Sigma$ we write $\xi^{i}=(\tau,\theta,\phi)$ which are related to the coordinates on $M$ by the following coordinate transformation
\begin{equation}
g_{ij}=\frac{\partial x^\alpha}{\partial \xi^i}\frac{\partial x^\beta}{\partial \xi^j}g_{\alpha \beta}. \label{10}
\end{equation}

Finally for the parametric equation on the induced metric $\Sigma$ we write
\begin{equation}
\Sigma: F(r,\tau)=r-a(\tau)=0.
\end{equation}

Note that in order to study the dynamics of the induced metric $\Sigma$, in the last equation we let the throat radius of the wormhole to be time dependent by incorporating the proper time on the shell i.e., $a=a(\tau)$. More specifically making use of the \eqref{10}, for the induced metric we have
\begin{equation}
\mathrm{d}s_{\Sigma}^{2}=-\mathrm{d}\tau^{2}+a(\tau)^{2}\left(\mathrm{d}\theta^{2}+\sin^{2}\theta\,\mathrm{d}\phi^{2}\right).\label{metric}
\end{equation}

The junction conditions on $\Sigma$ reads
\begin{equation}
{S^{i}}_{j}=-\frac{1}{8\pi}\left(\left[{K^{i}}_{j}\right]-{\delta^{i}}_{j}\,K\right).
\end{equation}

Note that in the last equation ${S^{i}}_{j}=diag(-\sigma,p_{\theta},p_{\phi})$
is the energy momentum tensor on the thin-shell, on the other hand $K$, and $[K_{ij}]$,
are defined as $K=trace\,[{K^{i}}_{i}]$ and $[K_{ij}]={K_{ij}}^{+}-{K_{ij}}^{-}$,
respectively.  Keeping this in mind, we can go on by writing the expression for the extrinsic curvature ${K^{i}}_{j}$ as follows
\begin{equation}
K_{ij}^{(\pm)}=-n_{\mu}^{(\pm)}\left(\frac{\partial^{2}x^{\mu}}{\partial\xi^{i}\partial\xi^{j}}+\Gamma_{\alpha\beta}^{\mu}\frac{\partial x^{\alpha}}{\partial\xi^{i}}\frac{\partial x^{\beta}}{\partial\xi^{j}}\right)_{\Sigma}.
\end{equation}

The unit vectors ${n_{\mu}}^{(\pm)}$, which are normal to $M^{(\pm)}$ are choosen as
\begin{equation}
n_{\mu}^{(\pm)}=\pm\left(\left\vert g^{\alpha\beta}\frac{\partial F}{\partial x^{\alpha}}\frac{\partial F}{\partial x^{\beta}}\right\vert ^{-1/2}\frac{\partial F}{\partial x^{\mu}}\right)_{\Sigma}.
\end{equation}

 Then the extrinsic curvature components are calculated 
as \cite{ern1}
\begin{equation}
K_{\theta}^{\theta \pm}=K_{\varphi}^{\varphi \pm}=\pm\frac{1}{a}\sqrt{f(a)+\dot{a}^{2}},\label{eq5}
\end{equation}
and 
\begin{equation}
K_{\tau}^ {\tau \pm}=\mp\frac{2\ddot{a}+f'(a)}{2\sqrt{f(a)+\dot{a}^{2}}},\label{eq6}
\end{equation}
in which the prime and the dot represent the derivatives with respect
to $r$ and $\tau$, respectively. Using the definitions $[K_{_{{i}{j}}}]\equiv K_{_{{i}{j}}}^{+}-K_{_{{i}{j}}}^{-}$,
and $K=tr[K_{{i}{j}}]=[K_{\;{i}}^{{i}}]$,
and the surface stress--energy tensor $S_{_{{i}{j}}}={\rm diag}(\sigma,p_{{\theta}},p_{{\varphi}})$
it follows the Lanczos equations on the shell 
\begin{equation}
-[K_{{i}{j}}]+Kg_{{i}{j}}=8\pi S_{{i}{j}}.\label{eq7}
\end{equation}

Note that for a given radius $a$, the energy density on the shell is
$\sigma$, while the pressure $p=p_{{\theta}}=p_{{\varphi}}$.
If we now combine the above results for the surface density 
\begin{equation}
\sigma=-\frac{\sqrt{f(a)+\dot{a}^{2}}}{2\pi a},\label{25}
\end{equation}
and the surface pressure 
\begin{equation}
p=\frac{\sqrt{f(a)+\dot{a}^{2}}}{8\pi}\left[\frac{2\ddot{a}+f'(a)}{f(a)+\dot{a}^{2}}+\frac{2}{a}\right].\label{26}
\end{equation}

Since we are going to study the wormhole stability at a static configuration we need to set $\dot{a}=0$, and $\ddot{a}=0$. For the surface density in static configuration it follows that
\begin{equation}
\sigma_{0}=-\frac{\sqrt{f(a_{0})}}{2\pi a_0},\label{27}
\end{equation}
and similary the surface pressure 
\begin{equation}
p_{0}=\frac{\sqrt{f(a_{0})}}{8\pi}\left[\frac{f'(a_{0})}{f(a_{0})}+\frac{2}{a_{0}}\right].\label{28}
\end{equation}

It's obvious from Eq. \eqref{27} that the surface density is negative, i.e. $\sigma_{0}<0$,
which implies that the weak and dominant energy conditions are violeted. At this stage it's interesting to calculate the amount of exotic matter at the wormhole by using the integral 
\begin{equation}
\Omega_{\sigma}=\int\sqrt{-g}\,\left(\rho+p_{r}\right)\,\mathrm{d}^{3}x.
\end{equation}

But since we are dealing with a thin--shell wormhole we must choose $p_{r}=0$ and $\rho=\sigma\delta(r-a)$,
where $\delta(r-a)$ is the Dirac delta function.  Solving this integral leads to the following result
\begin{equation}
\Omega_{\sigma}=\int_{0}^{2\pi}\int_{0}^{\pi}\int_{-\infty}^{\infty}\sigma\sqrt{-g}\,\delta(r-a)\mathrm{d}r\,\mathrm{d}\theta\,\mathrm{d}\phi.
\end{equation}

Finally the energy density of the exotic matter located on a thin shell surface is calculated as
\begin{equation}
\Omega_{\sigma}=-2a_0 \sqrt{1-\frac{r_0^4}{a_0^4}}.
\end{equation}

The wormhole can have attractive/repulsive nature, to see this let us calculate the 
the observes four--acceleration $a^{\mu}=u^{\nu}\nabla_{\nu}u^{\mu}$,
in which $u^{\mu}$ is the four velocity. We are left with the radial component of the acceleration
\begin{equation}
a^{r}=\Gamma_{tt}^{r}\left(\frac{\mathrm{d}t}{\mathrm{d}\tau}\right)^{2}=\frac{2 r_0^4}{a_0^5}. \label{accel}
\end{equation}

The test particle obeys the equation of motion if
\begin{equation}
\frac{\mathrm{d}^{2}r}{\mathrm{d}\tau^{2}}=-\Gamma_{tt}^{r}\left(\frac{\mathrm{d}t}{\mathrm{d}\tau}\right)^{2}=-a^{r}.
\end{equation}

From  Eq. \eqref{accel} follows three special cases. We can recover the geodesic equation if $a^{r}=0$, wormhole is attractive if $a^{r}>0$, and repulsive if $a^{r}<0$.

\section{Stability Analysis}

In this section we are going to analyze the stability of the CATSW. Starting from the energy conservation
it follows that 
\begin{equation}
\frac{d}{d\tau}\left(\sigma\mathcal{A}\right)+p\frac{d\mathcal{A}}{d\tau}=0,\label{34}
\end{equation}
where $\mathcal{A}=4\pi a^{2}$ is the area of the wormhole throat.
By replacing $\sigma(a)$ we can find the equation of motion as follows
\begin{equation}
\dot{a}^{2}=-V(a),\label{35}
\end{equation}
with the potential 
\begin{equation}
V(a)=f(a)-4\pi^{2} a^{2} \sigma^{2}(a).\label{36}
\end{equation}

In order to investigate the stability of CATSW let us expand the potential $V(a)$ around the static solution by writing 
\begin{equation}
V(a)=V(a_{0})+V^{\prime}(a_{0})(a-a_{0})+\frac{V^{\prime\prime}(a_{0})}{2}(a-a_{0})^{2}+O(a-a_{0})^{3}.\label{37}
\end{equation}
From Eq. (\ref{36}) we can find the first derivative of $V(a)$ 
\begin{equation}
V^{\prime}(a)=f^{\prime}(a)-8\pi^{2}a\,\sigma(a)\left[\sigma(a)+a \, \sigma^{\prime}(a) \right].\label{38}
\end{equation}

Furthermore if we use 
\begin{equation}
\sigma^{\prime}(a)=-\frac{2}{a}\left[\sigma(a)+p(a)\right],
\end{equation}
we can rewrite the first derivative of $V(a)$ as
\begin{equation}
V^{\prime}(a)=f^{\prime}(a)+8\pi^{2}a \,\sigma(a) \left[\sigma(a)+2p(a)\right].\label{39}
\end{equation}

The second derivative of the potential is 
\begin{equation}
V^{\prime\prime}(a) = f^{\prime\prime}(a)+8\pi^{2}\left\lbrace \left[a \sigma^{\prime}(a)+\sigma(a)\right]\left[\sigma(a)+2p(a)\right]+a\sigma(a)\left[\sigma^{\prime}(a)+2p^{\prime}(a)\right] \right\rbrace . \label{40}
\end{equation}

If we now use  $\sigma^{\prime}(a)+2p^{\prime}(a)=\sigma^{\prime}(a)[1+2p^{\prime}(a)/\sigma^{\prime}(a)]$,
and introduce the function  $p=\psi(\sigma)$, for the first derivative we have $\psi'=dp/d\sigma=p^{\prime}/\sigma^{\prime}$,
which implies $\sigma^{\prime}(a)+2p^{\prime}(a)=\sigma^{\prime}(a)(1+2\psi')$. 
Putting all together the above results, from Eq. \eqref{40} we obtain 
\begin{equation}
V^{\prime\prime}(a_{0})=f^{\prime\prime}(a_{0})-8\pi^{2}\left\lbrace \left[\sigma_{0}+2p_{0}\right]^{2}+2\sigma_{0}\left[\sigma_{0}+p_{0}\right](1+2\psi')\right\rbrace .\label{p10}
\end{equation}

The wormhole is stable if and only if $V^{\prime\prime}(a_{0})>0$. The equation of motion of the throat, for a small perturbation
becomes \cite{varela}%
\begin{equation}
\dot{a}^{2}+\frac{V^{\prime \prime }(a_{0})}{2}(a-a_{0})^{2}=0.
\end{equation}

Noted that for the condition of $V^{\prime \prime }(a_{0})\geq 0$, CATSW is stable where the motion of the throat is oscillatory with angular frequency $\omega =\sqrt{\frac{V^{\prime \prime }(a_{0})}{2}}$. In this work we are going to use five different models for the fluid to explore the stability analysis; a linear barotropic fluid(LBF) \cite{kuf2,varela}, chaplygin fluid (CF) \cite{cg1,cg2,GCCG,GCCG1}, logarithmic fluid (LogF) \cite{ali1,ali2}, polytropic eqiation of state for the fluid \cite{sarkar}, and finally Van der Waals Quintessence (VDWQ) \cite{Van}.

\subsection{Stability analysis of CATSW via the LBF}

In our first case, we choose a linear barotropic fluid with the equation of state given by \cite{kuf2,varela}
\begin{equation}
\psi=\omega\sigma,
\end{equation}
it follows that
\begin{equation}
\psi^{\prime}(\sigma_{0})=\omega.
\end{equation}

Note that $\omega$ is a constant parameter. In order to see more clearly the stability we show graphically the dependence of $\omega$ in terms of $a_{0}$ for different values of the parameter $r_0$ in Fig. 1. 

\begin{figure}[h!]
\includegraphics[width=0.30\textwidth]{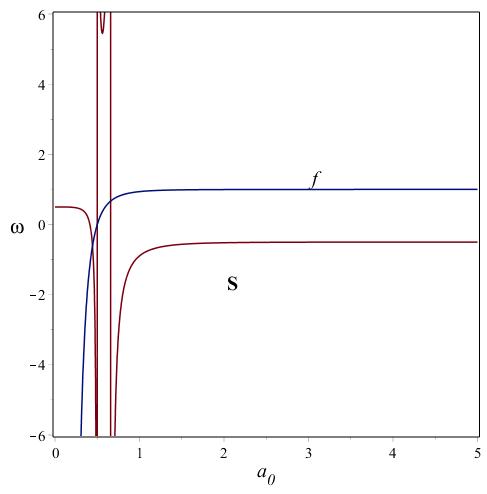} %
\includegraphics[width=0.30\textwidth]{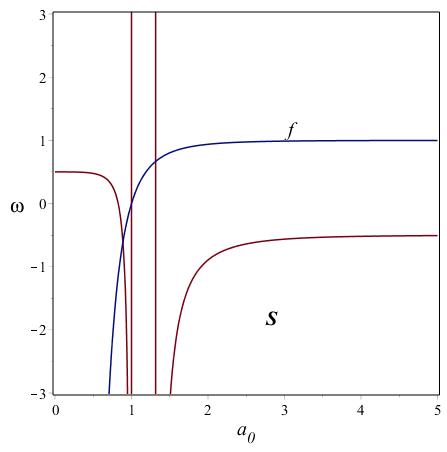} %
\includegraphics[width=0.30\textwidth]{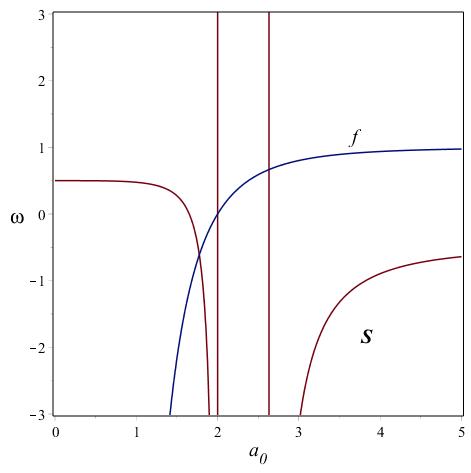} %
\caption{\small \textit{We plot the stability regions as a function of $\omega$ and radius of the throat $a_{0}$. We have choosen three different values  $r_0=0.5, r_0=1$ and $r_0=2$.} }
\end{figure}

\subsection{Stability analysis of CATSW via the CF}

According to the chaplygin fluid (CF), we can model the fluid with the following equation of state \cite{cg1,cg2,GCCG,GCCG1}
\begin{equation}
\psi=\omega \left(\frac{1}{\sigma}-\frac{1}{\sigma_0}\right)+p_0,
\end{equation}
to find
\begin{equation}
\psi^{\prime}(\sigma_{0})=-\frac{\omega}{\sigma_0^{2}}.
\end{equation}

To see the stabiliy regions let us show graphically the dependence of $\omega$ in terms of $a_{0}$ for different values of the parameter $r_0$, given in Fig. 2.

\begin{figure}[h!]
\includegraphics[width=0.30\textwidth]{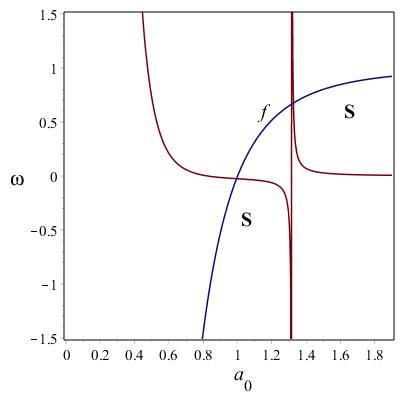} %
\includegraphics[width=0.30\textwidth]{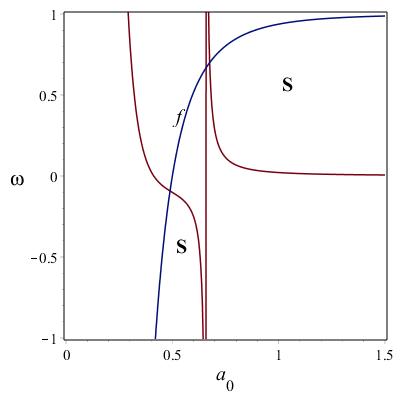}
\includegraphics[width=0.30\textwidth]{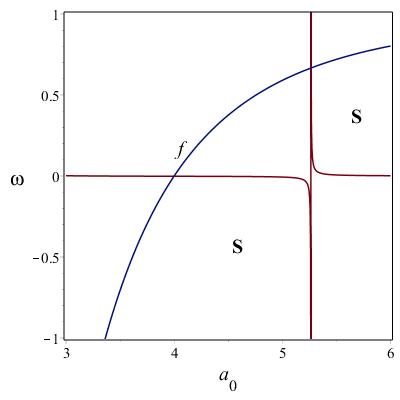} %
\caption{\small \textit{Here we plot the stability regions as a function of $\omega$ and radius of the throat $a_{0}$. We have choosen three different values  $r_0=1, r_0=0.5$ and $r_0=4$. } }
\end{figure}

\subsection{Stability analysis of CATSW via the LBF}

Our next example is the logarithmic fluid (LogF) \cite{ali1,ali2}, with the equation of state 
\begin{equation}
\psi=\omega \ln \left(\frac{\sigma}{\sigma_0}\right)+p_0,
\end{equation}
then
\begin{equation}
\psi^{\prime}(\sigma_{0})=\frac{\omega}{\sigma_0}.
\end{equation}

For detailed information we can show graphically the dependence of $\omega$ in terms of $a_{0}$ by choosing different values of the parameter $r_0$, in Fig. 3. 

\begin{figure}[h!]
\includegraphics[width=0.33\textwidth]{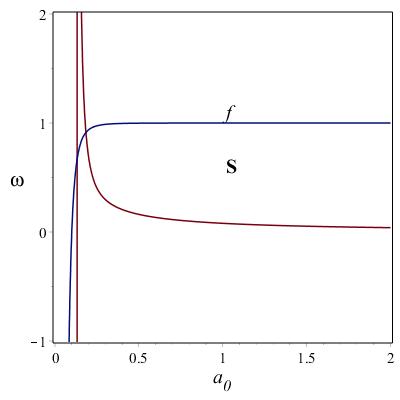} %
\includegraphics[width=0.33\textwidth]{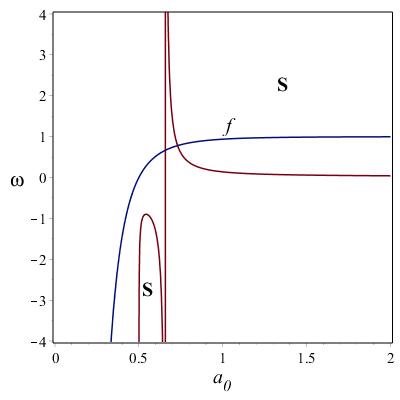} %
\includegraphics[width=0.33\textwidth]{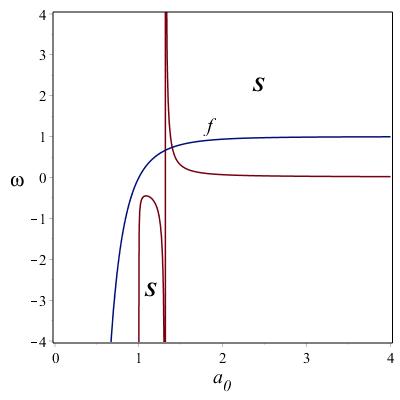} %
\includegraphics[width=0.33\textwidth]{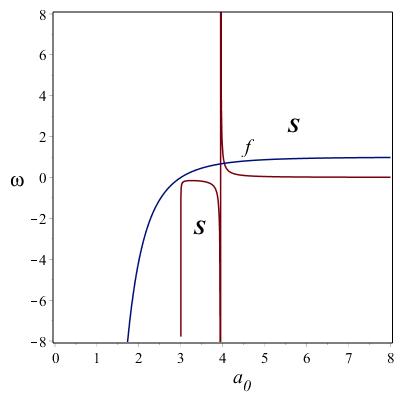}
\caption{\small \textit{The stability regions as a function of $\omega$ and radius of the throat $a_{0}$, in which we have choosen three different values $r_{0}=0.1, r_0=0.5, r_0=1$ and $r_0=3$.} }
\end{figure}

\subsection{Stability analysis of CATSW via polytropic fluid}

The equation of state for the fluid according to the polytropic model can be written as \cite{sarkar,poli}
\begin{equation}
\psi=\omega \sigma^{\gamma},
\end{equation}

It follows that
\begin{equation}
\psi^{\prime}(\sigma_{0})=\omega\,\gamma \, \sigma_0^{\gamma -1}
\end{equation}

For detailed information we plot $\omega$ in terms of $a_{0}$ by choosing different values of the parameter $r_0$, as shown in Fig. 4. 

\begin{figure}[h!]
\includegraphics[width=0.30\textwidth]{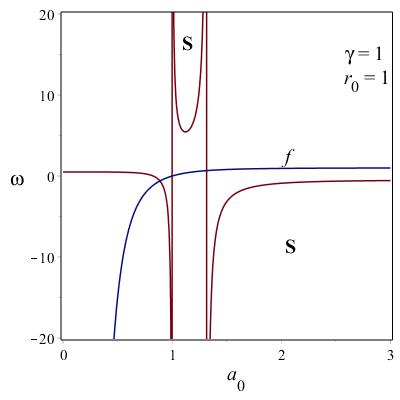} %
\includegraphics[width=0.30\textwidth]{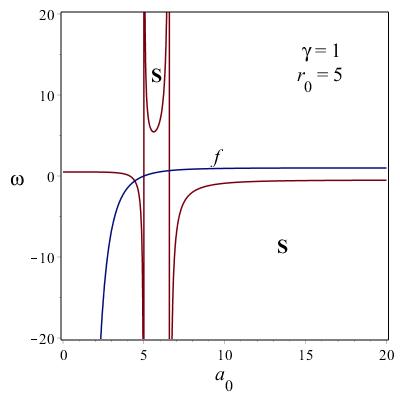} %
\includegraphics[width=0.30\textwidth]{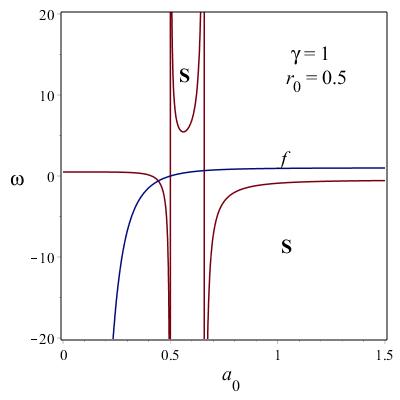} %
\caption{\small \textit{Stability regions in terms of $\omega$ and radius of the throat $a_{0}$ for different values of $r_0$.} }
\end{figure}

\subsection{Stability analysis of CATSW via Van der Waals Quintessence}

The equation of state for the fluid according to the Van der Waals quintessence model can be written as \cite{Van}
\begin{equation}
\psi=\frac{\gamma\sigma}{1-B\sigma}-\alpha \sigma^2 
\end{equation}
where $\gamma$, B and $\alpha$ are constants.
It follows that
\begin{equation}
\psi^{\prime}(\sigma_{0})=\frac{-2\alpha\sigma(B\sigma-1)^2+\gamma}{(B\alpha-1)^2}
\end{equation}

For more useful informations for the acustic wormhole stability we plot $\alpha$ and $a_0$ for different values of the parameters $\gamma$, $r_0$ and $B$, as shown in Fig. 5.

\begin{figure}[h!]
\includegraphics[width=0.33\textwidth]{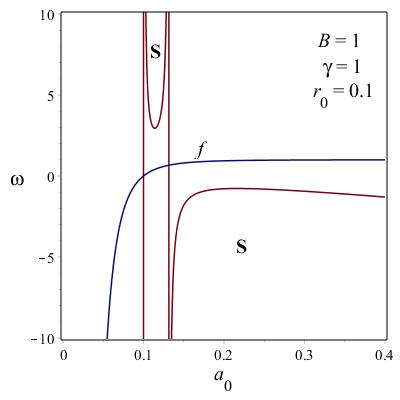} %
\includegraphics[width=0.33\textwidth]{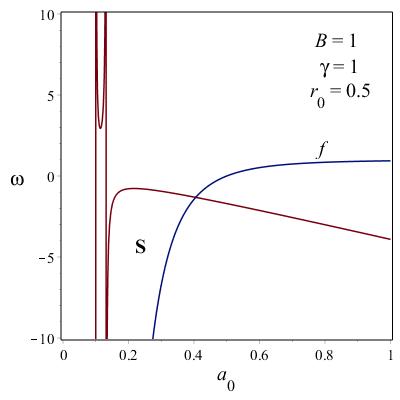} %
\includegraphics[width=0.33\textwidth]{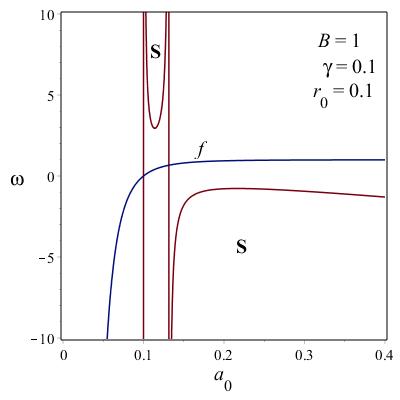} %
\includegraphics[width=0.33\textwidth]{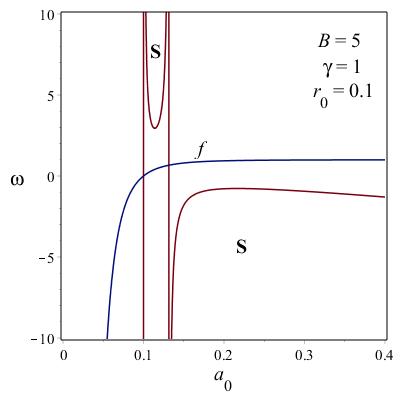}
\caption{\small \textit{Stability regions of CATSW  in terms of $\omega$ and radius of the throat $a_{0}$ for  different values of $\gamma$, $B$, and $r_0$.} }
\end{figure}

\section{Conclusion}

In this paper we constructed a spherically symmetric, canonical acoustic thin shell wormhole, in the context of analogue gravity systems. We have used cut and paste method to join together two regular regions, then, we have computed the analogue surface density and surface pressure of the fluid.  The stability analyses is carried out using a  linear barotropic fluid, chaplygin fluid, logarithmic fluid, polytropic fluid, Van der Waals Quintessence, for the fluid and shown that the wormhole can be stabile if one chooses suitable parameter values. We show that acoustic fluid with negative energy is required at the throat to keep the wormhole stable.
\section{acknowledgement}
We wish to thank the anonymous referees for valuable suggestions.


\begin{thebibliography}{10}


\bibitem{fl} L. Flamm, Phys. Z. 17, 448 (1916).

\bibitem{er} A. Einstein and N. Rosen, Phys. Rev. 48, 73-77 (1935).

\bibitem{mth1} M. S. Morris and K. S. Thorne,  Am. J. Phys. 56, 395 (1988).


\bibitem{mth2}  M. Morris, K. S. Thorne and U. Yurtsever,  Phys.Rev. Lett. 61, 1446 (1988).

\bibitem{mv1} M. Visser, Lorentzian Wormholes (AIP Press, New York, 1996).

\bibitem{mv2} M.Visser , Nuclear Physics B 328, 203 (1989).

\bibitem{mv3} M.Visser,  Phys. Rev. D, 39, 3182 (1989).

\bibitem{is} W. Israel,  Nuovo Cimento, 44B, 1 (1966).


\bibitem{eiroa1} E. F. Eiroa and G. E. Romero, Gen. Relativ. Gravit. 36, 651 (2004).

\bibitem{ayan1} Ayan Banerjee, 	Int J Theor Phys, 2013, Volume 52, Issue 8, pp 2943-2958.

\bibitem{lobo1} F. S. N. Lobo and P. Crawford, Class. Quantum Grav. 21, 391 (2004).

\bibitem{eiroa2} Ernesto F. Eiroa, Claudio Simeone, Phys.Rev. D 71 (2005) 127501.

\bibitem{ali1} M. Halilsoy, A. Ovgun, S. Habib Mazharimousavi, Eur. Phys. J. C (2014) 74:2796

\bibitem{rah} F. Rahaman, M. Kalam and  S. Chakraborty, Gen. Rel. Grav. 38, 1687-1695 (2006).

\bibitem{mazh1} S. Habib Mazharimousavi, M. Halilsoy, 	Eur. Phys. J. C (2014) 74:3073

\bibitem{ali} A. Ovgun, Eur. Phys. J. Plus (2016) 131: 389

\bibitem{kimet} Kimet Jusufi, Eur. Phys. J. C (2016) 76:608 



\bibitem{po} E. Poisson and M. Visser,  Phys. Rev. D 52, 7318 (1995).

\bibitem{eiroa3} E. F. Eiroa and C. Simeone, Phys.Rev.D 71 127501 (2005).

\bibitem{ali2}  A. Ovgun, I. Sakalli, TMPh, 2017, V. 190, No. 1

\bibitem{ern1} N. M. Garcia, F.S.N. Lobo and M. Visser, Phys. Rev. D 86, 044026 (2012).
\bibitem{ern2} Ernesto F. Eiroa Phys.Rev.D 80, 044033 (2009).

\bibitem{lemos1} J.P. S. Lemos and F. S. N. Lobo, Phys.Rev. D 69, 104007 (2004).

\bibitem{lemos2} J.P. S. Lemos and F. S. N. Lobo, Phys.Rev.D 78, 044030 (2008).

\bibitem{ayan2} Farook Rahaman, A. Banerjee, I. Radinschi,  Int J Theor Phys (2012) 51: 1680. 

\bibitem{myrzakulov} Ratbay Myrzakulov, Lorenzo Sebastiani, Sunny Vagnozzi, Sergio Zerbini, Class. Quant. Grav. \textbf{33} (2016) 12, 125005

\bibitem{kim} Kim S W, Lee H, Kim S K and Yang J 1993 Phys. Lett. A \textbf{183}, 359

\bibitem{peter1} Peter K. F. Kuhfittig, Acta Phys. Polon. B \textbf{41}, 2017-2019, 2010

\bibitem{sharifcy1} M. Sharif and M. Azam, JCAP \textbf{04}, 023 (2013).


\bibitem{sharifcy2} M. Sharif, M. Azam, Eur. Phys. J. C \textbf{73}, 2407, 2013.

\bibitem{sh1} M. Sharif, S. Mumtaz, Astrophys.Space Sci. 361, no.7, 218 (2016).

\bibitem{sh2} M. Sharif, F. Javed, Gen.Rel.Grav. 48, no.12, 158 (2016).

\bibitem{sh3}  E. F. Eiroa, G. F. Aguirre, Phys.Rev. D94, no.4, 044016 (2016).


\bibitem{unruh1} W.G. Unruh, Phys. Rev. Lett. 46, 1351 (1981).

\bibitem{Jeff} J. Steinhauer, Nature Phys., \textbf{10}, 864 (2014); J.
Steinhauer, Phys. Rev. D \textbf{92}, 024043 (2015).

\bibitem{visser1} M. Visser, Class. Quantum Grav. 15, 1767 (1998).

\bibitem{unruh2} R. Schutzhold and W. G. Unruh, Physical Review D 66, 044019 (2002)

\bibitem{visser2} M. Visser and S. Weinfurtner, Classical and Quantum Gravity 22, 2493, (2005)

\bibitem{saavedra1}S. Lepe and J. Saavedra, 	Phys. Lett. B 617, 174-181 (2005).

\bibitem{saavedra2}J. Saavedra, Mod. Phys. Lett. A 21, 1601-1608 (2006).

\bibitem{lib} Ramit Dey, Stefano Liberati, Rodrigo Turcati, Phys. Rev. D 94, 104068 (2016)

\bibitem{vieira} H. S. Vieira, V. B. Bezerra, Gen Relativ Gravit (2016) 48:88

\bibitem{ramon} Ramon Becar, Pablo Gonzalez, Gustavo Pulgar, Joel Saavedra, Int.J.Mod.Phys. A, 25 (2010) 1463-1475

\bibitem{nandi} Kamal Kanti Nandi, Yuan-Zhong Zhang, Rong-Gen Cai, \url{gr-qc/0409085v5 }

\bibitem{carlos} Carlos Barcelo, Stefano Liberati, Matt Visser, 	Living Rev. Rel. 8, 12, 2005

\bibitem{kuf2} Kuhfittig P.K.F., Ann. Phys. 355, 115 (2015).

\bibitem{varela} Varela V., Phys. Rev. D. 92, 044002 (2015).

\bibitem{cg1} Eiroa E.F., Simeone C., Phys. Rev. D 76, 024021 (2007).
 
\bibitem{cg2} Lobo F.S.N., Phys. Rev. D 73, 064028 (2006).


\bibitem{sarkar} Sanjay Sarkar, Int J Theor Phys, 55, Issue 1, pp 481-494.

\bibitem{poli} Mubasher Jamil, Peter K.F. Kuhfittig, Farook Rahaman, Sk. A Rakib, 	Eur. Phys. J. C, 67:513-520, (2010).
\bibitem{Van} M. Sharif and S. Mumtaz, Advances in High Energy Physics, Volume 2016, Article ID 2868750 (2016).

\bibitem{GCCG} M. Azam, Astrophys Space Sci 361, 96 (2016).




\bibitem{GCCG1}P.F. Gonzaalez-Diaz, Phys. Rev. D 68, 021303 (2003).





\end{thebibliography}
\end{document}